\documentclass[11pt]{article}
\usepackage{epsfig}
\usepackage{fullpage}
\usepackage{graphics}
\usepackage{amsmath}
\usepackage{amsfonts}
\newcommand{\qed}{{\hspace*{\fill}}}
\def\proof{{\noindent{\em Proof : }}}

\begin{document} 
 
\newtheorem{thm}{Theorem} 
\newtheorem{lemma}[thm]{Lemma} 
\newtheorem{cor}[thm]{Corollary} 
\newtheorem{con}[thm]{Conjecture} 
\newtheorem{prop}[thm]{Proposition} 
\newtheorem{alg}[thm]{Algorithm} 
\def\ZZ{{\mathbb Z}} 
\def\RR{{\mathbb R}} 
\def\NN{{\mathbb N}} 
\def\CC{{\mathbb C}} 
\def\II{{\mathbb I}} 
\def\intcol{{\mbox{intcol}}} 
 
\title{Decomposing Finite Abelian Groups} 
\author{Kevin K. H. Cheung\thanks{
Ph.D. student, Department of Combinatorics and Optimization,
Faculty of Mathematics, University of Waterloo, Waterloo, Ontario,
N2L 3G1, Canada.  E-mail: kkhcheun@math.uwaterloo.ca. 
Research of this author was supported by NSERC PGSB}
\and
Michele Mosca\thanks{Department of Combinatorics and Optimization,
Faculty of Mathematics, University of Waterloo, Waterloo, Ontario,
N2L 3G1, Canada.  E-mail: mmosca@cacr.math.uwaterloo.ca.  
Research of this author was supported by NSERC}
}
 
\maketitle 
 
\begin{abstract} 
This paper describes a quantum algorithm for efficiently 
decomposing finite Abelian groups.  Such a decomposition is needed 
in order to apply the Abelian hidden subgroup algorithm. Such a 
decomposition (assuming the Generalized Riemann Hypothesis) also 
leads to an efficient algorithm for computing class numbers (known 
to be at least as difficult as factoring). 
\end{abstract} 
 
\section{Introduction} 
 
The work by Shor~\cite{Sh94} on factoring and finding discrete 
logarithms over $\ZZ_n^*$ can be generalized to solve the Abelian 
Hidden Subgroup Problem (see for example \cite{Vaz97}, 
\cite{Hoy97}, \cite{ME} ).  These algorithms find the hidden 
subgroup of a function $f: G \rightarrow S$, where $G = 
\ZZ_{N_1}\times 
\cdot\cdot\cdot \ZZ_{N_l}$, for some integers $N_1, N_2, \ldots, N_l$. 
Any Abelian group $G$ is isomorphic to such a product of cyclic 
groups, however it is not always known how to find such an 
isomorphism efficiently.  Consider for example the group 
$\ZZ_{N}^*$, the multiplicative group of integers modulo $N$. This 
is an Abelian group we can compute in efficiently, yet no known 
classical algorithm can efficiently find its decomposition into a 
product of cyclic groups. Consider also the class group of a 
quadratic number field. This group is also Abelian, and finding its 
decomposition into a product of finite cyclic groups will give us 
the size of the group and therefore the class number of the 
quadratic number field.  As Watrous \cite{Wat00} points out, 
assuming the generalized Riemann Hypothesis we can apply the 
algorithm in this paper and efficiently find class numbers (a 
problem known to be at least as hard as factoring). 
 
In this paper, we show how we can make use of the solution to the 
Abelian Hidden Subgroup Problem to decompose a finite Abelian 
group.  Such decompositions makes it possible to apply the Abelian 
Hidden Subgroup algorithm to a larger class of Abelian groups. 
 
\section{Integer Arithmetic Basics} 
 
A nonsingular integral matrix $U$ is called \textit{unimodular} if $U$ is 
has determinant $\pm 1$.  It is easy to check that $U$ is unimodular if 
and only if $U^{-1}$ is unimodular. 
 
 
 
The following operations on a matrix are called 
\textit{elementary (unimodular) column operations}: 
1. exchanging two columns; 2. multiplying a column by -1; 
3. adding an integral multiple of one column to another column. 
 
We define \textit{elementary row operations} similarly. 
 
 
 
\begin{thm}\label{SNF} 
For any integral matrix $A$, one can find in polynomial time using 
elementary row and column operations 
unimodular matrices $U$ and 
$V$ such that $UAV=\left[\begin{array}{cc} D&0\\ 0&0 \end{array}\right]$ 
where $D=Diag(d_1,...,d_k)$ with positive integers such that 
$d_1|d_2|...|d_k$ and for each $i$, the product 
$d_1...d_i$ is equal to the g.c.d. of the subdeterminants of $A$ 
of order $i$. 
We call the matrix $\left[\begin{array}{cc} D&0\\ 0&0 \end{array}\right]$ 
the Smith normal form (abbreviated as SNF) of $A$. 
\end{thm} 
\proof See Kannan and Bachem~\cite{KB}. \qed
 
\section{Group Theory Basics} 
 
Recall that a group $G$ is said to be \textit{Abelian} if for all $a,b\in G$, 
$a\cdot b = b\cdot a$.  In this paper, all groups are finite 
Abelian unless otherwise stated. 
$G$ is said to be \textit{cyclic} if there exists $a\in G$, 
such that $G=\{a^n|n\in \ZZ\}$.  Here, we call $a$ a 
\textit{generator} of $G$. 
$H \subseteq G$ is called a \textit{subgroup} of $G$ if $H$ is a group 
under the operation induced by $G$.  In this case, we write $H \leq G$. 
Let $a\in G$.  The set $aH =\{g\in G|g=ah$ for some $h\in H\}$ is 
called a \textit{coset} of $H$ in $G$ determined by $a$. 
 
Let $a\in G$.  If $a^n=e$ for some $n\in \NN$, then $a$ is said to have 
\textit{finite order}.  The smallest such $n$ is called the $order$ of 
$a$, denoted by \textit{ord}$(a)$.  It is easy to see that the elements in 
$\{e, a, a^2,...,a^{n-1}\}$ form a subgroup.  We call this subgroup the 
\textit{the cyclic subgroup generated by} 
$a$ and we denote it by $\langle a\rangle$. 
 
Let $G_1,G_2$ be groups such 
that $G_1\cap G_2 = \{e\}$. 
The set $\{a_1a_2 \mid a_1\in G_1$, $a_2\in G_2\}$, 
denoted by $G_1 \oplus G_2$, is called the \textit{direct sum} of $G_1$ and 
$G_2$.  Note that $G_1 \oplus G_2$ is a group under binary operation 
$\cdot$ such that $(a_1b_1)(a_2b_2)=(a_1b_1)(a_2b_2)$. 
 
Let $G$ be a group and $p$ be a prime number.  Let $P\leq G$. 
Then $P$ is called a \textit{Sylow p-subgroup} of $G$ if $|P| = p^{\alpha}$ 
for some $\alpha \in \NN$ such that $p^{\alpha}$ divides $|G|$ but 
$p^{\alpha+1}$ does not. 
 
We first quote a few classical results without proof.  The interested 
reader can refer to a standard text on group theory.
 
 
\begin{thm}\label{normal} 
If $N$ is a subgroup of an Abelian group $G$, then the set of 
cosets of $N$ forms a group under the coset multiplication given 
by 
\begin{equation*} 
  aNbN= abN 
\end{equation*} 
for all $a,b\in G$.  The group is denoted by $G/N$. 
\end{thm} 
 
\begin{thm}\label{generate} 
Let $N$ be a subgroup of a finite Abelian group $G$. 
If $a_1,...,a_k$ generates $G$, then 
$a_1N,...,a_kN$ generates $G/N$. 
\end{thm} 
 
\begin{thm}\label{product} 
  A finite Abelian group can be expressed as a direct sum of its 
  Sylow $p$-subgroups. 
\end{thm} 
%
 
 
\begin{thm}\label{subgroup} 
Let $K$ be a subgroup of $G=G_{p_1} \oplus \cdot\cdot\cdot \oplus G_{p_l}$ 
where $G_{p_i}$ is a Sylow $p_i$-subgroup for $i=1,...,l$ and 
$p_1,...,p_l$ are distinct primes. 
Then there exists $K_{p_i} \leq G_{p_i}$, $i=1,...,l$, such that 
$K=K_{p_1} \oplus \cdot\cdot\cdot \oplus K_{p_l}$. 
\end{thm} 
%
%
 
The next theorem is an important result on finite Abelian groups. 
 
\begin{thm}\label{fund} 
  (Fundamental Theorem of Finite Abelian Groups). 
  Any finite Abelian group can be decomposed as a direct sum of cyclic 
  subgroups of prime power order. 
\end{thm} 
 
In this paper, we shall give an algorithm to find the decomposition.
 
The next theorem is an integral part of the algorithm. 
 
\begin{thm}~\label{reduce} 
Given 
a generating set  $\{a_1,...,a_k\}$ of a finite 
Abelian group $G$ and a matrix $M$ such that 
$a_1^{x_1}\cdot\cdot\cdot a_k^{x_k} = e$ 
if and only if $\mathbf{x}$ $= (x_1,...,x_k)^T\in$ 
intcol($M$) where intcol($M$) denotes the set of vectors 
obtainable by taking integer linear combinations of 
columns of $M$, 
we can find in polynomial time (in the size of $M$) 
$g_1,...,g_l$ with $l\leq k$ such that 
$G = \langle g_1\rangle\oplus\cdot\cdot\cdot \oplus \langle g_l\rangle$. 
\end{thm} 
 
\proof (Adapted from Algorithm 4.1.3 in \cite{cohen}.) 
By Theorem~\ref{SNF}, we can find in polynomial time unimodular matrices 
$U$ and $V$ such that 
$U^{-1}MV = \left[ 
  \begin{array}{cc} 
    D & 0 \\ 
    0 & 0 
  \end{array} 
  \right]$ 
where $D$ is a diagonal matrix with 
diagonal entries $d_1,...,d_m$. 
Since $V$ is unimodular, intcol($MV$)=intcol($M$). 
Thus, 
$a_1^{x_1}\cdot\cdot\cdot a_k^{x_k} = e$ 
if and only if $\mathbf{x}$ $= (x_1,...,x_k)^T\in$ 
intcol($MV$). 
For each $i=1,...,k$, 
set $a_i'= a_1^{U_{1i}}\cdot\cdot\cdot a_k^{U_{ki}}$. 
Then 
\begin{eqnarray*} 
 & &{a_1'}^{x_1}\cdot\cdot\cdot {a_k'}^{x_k} = e \\ 
 &\Leftrightarrow & (x_1,...,x_k)^T \in \mbox{intcol}(U^{-1}MV) \\ 
 &\Leftrightarrow & d_i|x_i \mbox{ for } i=1,...,m, \mbox{ and } 
   x_i = 0 \mbox{ for } i=m+1,...,k 
\end{eqnarray*} 
Since $G$ is finite, we must have $m=k$.  Otherwise, $G$ will 
have an element of infinite order. 
Let $j$ be the smallest index such that $d_j > 1$. 
Set $g_i = a_{i+j-1}'$ for $i=1,..,l$ where 
$l= m-j+1$. 
It is clear that $g_1,...,g_l$ still generate $G$ and 
$g_i$ has order $d_{i+j-1}$ for $i=1,...,l$. 
Therefore, if $0 \leq x_i <$ ord($g_i$), then 
${g_1}^{x_1}\cdot\cdot\cdot {g_l}^{x_l} = e$ 
implies that $x_i=0$ for all $i$. 
Hence 
$G = \langle g_1\rangle\oplus\cdot\cdot\cdot \oplus \langle g_l\rangle$. 
\qed
 
\section{Hidden Subgroup Problem} 
 
Let $G = \ZZ_{N_1}\times \cdot\cdot\cdot \ZZ_{N_l}$ where 
the $N_j$, $j=1,...,l$ are prime powers. 
We are given $f:G \rightarrow S$ for some finite set $S$ 
that is constant on cosets of some $K \leq G$ but distinct on 
each coset.  (The case when 
distinct cosets are not mapped to distinct elements is addressed 
in Boneh and Lipton~\cite{BL} and in the Appendix of \cite{ME}. 
Here, we need $m< |K|$ where $m$ is the maximum number of cosets 
that get mapped to the same output.) 
The hidden subgroup $K$ is 
\begin{equation*} 
\{k\in G \mid f(x) = f(x+k) \mbox{ for all } x\in G\}. 
\end{equation*} 
The Hidden Subgroup Problem is to find generators for $K$ given 
only $f$ and $G$. 
 
There exist polynomial-time quantum algorithms to solve this problem. 
 
\begin{cor}\label{order} 
  Let $a$ be an element of a group $G$.  The order $r$ of $a$ 
  can be found in random quantum polynomial time. 
\end{cor} 
\proof
Consider the function $f$ from $\ZZ$ to the group $G$ where 
$f(x)=a^x$.  Then $f(x)=f(y)$ if and only if $x-y\in r\ZZ$. 
The hidden subgroup is $K=r\ZZ$ and a generator for $K$ gives us the 
order $r$ of $a$. \qed
 
Using Corollary~\ref{order}, one can deduce the result by 
Shor~\cite{Sh94}. 
 
\begin{thm}\label{shor} 
  Factoring can be solved in random quantum polynomial time. 
\end{thm} 
 
In the next section, we shall show how to use the algorithm 
for finding hidden subgroup to decompose finite Abelian groups. 
 
\section{Decomposing Abelian Groups} 
 
By Theorem~\ref{fund}, we know that we can decompose a finite 
Abelian group into a direct sum of cyclic groups of prime power 
order. This problem was discussed briefly in \cite{mike}. We make 
four assumptions on the group $G$: 
 
\begin{enumerate} 
\item We have a unique binary representation for each element of $G$ 
and we can efficiently recognize if a binary string represents an 
element of $G$ or not. 
 
\item Using the binary representation, for any $a\in G$, we can efficiently 
construct a quantum network for implementing 
$U_a:|y\rangle \rightarrow |ay\rangle$. 
 
\item We can efficiently find a generating set for $G$. 
 
\item The orders of the generators are of prime power order. 
\end{enumerate} 
 
To meet the third assumption, it suffices to have an upper bound of 
$2^k$ on the size of the groups we work with for some 
$k \in \Theta(\log |G|)$ and 
that we can efficiently sample elements of $G$ uniformly at random. 
(If we do not have such a bound, we can easily devise a procedure
that tries an increasing sequence of values for $k$ and still has
expected running time in $O$(poly $\log |G|$)).
Let $K$ be a proper subgroup of $G$.  Then there are at least two 
cosets of $K$.  If we randomly sample an element $x$ from $G$, then 
with probability at least $1/2$, the subgroup spanned by $x$ and 
$K$ will have size at least twice that of $K$ because the elements 
$xk$ for all $k\in K$ are in the span. Hence, it takes an expected 
number of at most $(1/(1/2))k=2k$ samples to obtain a generating 
set $\mathcal{G}$ for $G$ and therefore $2k + c \sqrt{k}$ samples 
will find a generating set with probability in $1 - \epsilon^{c}$
for some $\epsilon \in (0,1)$ (by a Chernoff bound.)
 
Now we may assume that the order of the elements are of a 
prime power.  Let $a$ be an element in $\mathcal{G}$ 
with order $pq$ where $(p,q)=1$, $p \neq 1$ and $q \neq 1$. 
Note that $p$ and $q$ can be determined efficiently as a result 
of Corollary~\ref{order} and Theorem~\ref{shor}. 
By the Euclidean algorithm, we can find 
$r,s$ such that $rp+sq=1$.  Thus $(a^p)^r(a^q)^s=a$. 
Hence, replacing $a$ with $a^q$ and $a^p$ still leaves us 
with a generating set.  We repeat this procedure until 
each element in $\mathcal{G}$ has prime power order. 
 
Since we know the order $pq$ of $a$, we can efficiently compute 
$a^{-1} 
= a^{pq-1}$ and therefore efficiently perform 
the necessary uncomputation in order to satisfy 
the second assumption. 
 
By Theorem~\ref{product}, we have 
$G = G_{p_1} \oplus \cdot\cdot\cdot \oplus G_{p_l}$ where 
$p_i$ is a prime for all $i=1,...,l$ and $G_{p_i}$ 
is a Sylow $p_i$-subgroup of $G$. 
Let $S_j$ be the set of 
all the elements in $\mathcal{G}$ having 
order a power of the prime $p_j$.  For $a \in S_j$, let $K_a$ denote 
the (cyclic) subgroup generated by $a$. By Theorem~\ref{subgroup}, 
we have $K_a = K_{p_1} \oplus \cdot\cdot\cdot \oplus K_{p_l}$ where 
$K_{p_i} \leq G_{p_i}$ for all $i=1,...,l$.  Since $|K_a|$ is a 
power of $p_j$, we must have $K_a \leq G_{p_j}$. 
Thus $S_j \in G_{p_j}$.  Since $\mathcal{G}$ generates $G$, 
$S_j$ generates $G_{p_j}$.  Hence, we can first find the decomposition 
for each of the Sylow $p$-subgroups of $G$ and then take their 
product to obtain a decomposition of $G$. 
 
There are two primary reasons why we want to have the fourth 
assumption.  One reason is that we want to minimize the amount of 
quantum computing resources required in any implementation.  It is 
therefore advisable to decompose the problem whenever it is 
possible.  The second reason is that working with $p$-groups can 
greatly simplify the amount of algebra one needs to perform to 
recover the generators.  This latter point will be elaborated at 
the end of the section.  In the meantime, we present the algorithm 
given in \cite{mike} which finds generators of a group $G$ with 
prime power order. 
 
\begin{alg} Decompose\_Group($a_1,...,a_k$) 
 
\textbf{Input:} 
\begin{itemize} 
\item A generating set $\{a_1,...,a_k\}$ of the group $G$. 
\item The maximum order $q=p^r$ of the elements $a_1,...,a_k$. 
\end{itemize} 
 
\textbf{Output:} 
\begin{itemize} 
\item A set of elements $g_1,...,g_l$, $l\leq k$, from the group $G$. 
\end{itemize} 
 
\textbf{Procedure:} 
\begin{enumerate} 
\item Define $g:\ZZ_q^k\rightarrow G$ by mapping 
$(x_1,...,x_k)\rightarrow g(\mathbf{x}) = 
a_1^{x_1}\cdot\cdot\cdot a_k^{x_k}$. 
Find generators for the hidden subgroup $K$ of 
$\ZZ_q^k$ as defined by the function $g$. 
\item Compute a set $\mathbf{y_1},...,\mathbf{y_l}$ $\in \ZZ_q^k/K$ 
 of generators for $\ZZ_q^k/K$. 
\item Output $\{g(\mathbf{y_1}),...,g(\mathbf{y_l})\}$. 
\end{enumerate} 
\end{alg} 
 
To see the correctness of this algorithm, 
observe that the hidden subgroup $K$ is the 
set $\{ (x_1,...,x_k) \mid~ a_1^{x_1}\cdot\cdot\cdot a_k^{x_k} = e \}$. 
We therefore have an isomorphism between $\ZZ_q^k/K$ and $G$.  If 
$\mathbf{y_1},...,\mathbf{y_l}$ are generators for $\ZZ_q^k/K$, 
then 
$\{g(\mathbf{y_1}),...,g(\mathbf{y_l})\}$ are generators for $G$. 
 
We now elaborate on how it is possible to find 
generators for $\ZZ_q^k/K$. 
Observe that $e_1,...,e_k$ generate $\ZZ_q^k$ where 
$e_i$ is a 0,1-vector with a 1 in the $i$th co-ordinate. 
Further, if we let $M=q\II$ where $\II$ is the $k\times k$ identity 
matrix, then
$x_1e_1+...+x_ke_k=\mathbf{0}$ (in $\ZZ_q^k$) if and only if 
$\mathbf{x}$ $\in \intcol(M)$. 
By Theorem~\ref{generate}, 
$e_1+K,...,e_k+K$ generate $\ZZ_q^k/K$. 
Note that 
\begin{equation*} 
  v_1(e_1+K)+...+v_k(e_k+K) = K 
\end{equation*} 
if and only if $\II\mathbf{v}$ $\in K$ 
where $\II$ is the matrix $[e_1...e_k]$ 
and $\mathbf{v}$ $=(v_1,...,v_k)^T$. 
Let $A$ be the matrix the columns of 
which generate $K$. 
Note that $\II\mathbf{v}$ $\in K$  if and only if there 
exists a vector $\mathbf{x}$ such that 
\begin{eqnarray*} 
 & & \II\mathbf{v} = A\mathbf{x}  \\ 
  &\Leftrightarrow & \II\mathbf{v} = \II A\mathbf{x} \\ 
  &\Leftrightarrow & \II(\mathbf{v}-A\mathbf{x}) = \mathbf{0} \\ 
  &\Leftrightarrow & \mathbf{v}-A\mathbf{x} \in \intcol(M) \\ 
  &\Leftrightarrow & \mathbf{v} \in \intcol([M|A]). 
\end{eqnarray*} 
Applying Theorem~\ref{reduce} to $\{e_i+K \mid i=1,...,k\}$ and 
$M'=[M|A]$, we obtain $\mathbf{y_1},...,\mathbf{y_l}$ 
$\in \ZZ_q^k/K$ 
such that 
$\ZZ_q^k/K=\langle y_1\rangle\oplus\cdot\cdot\cdot\oplus \langle y_l\rangle$ 
as desired.

Technically, we need not work with each Sylow $p$-subgroup of $G$ 
separately. Suppose $a_{i1},...,a_{ik_i}$ generates the group 
$G_{p_i}$. Let $q_i=p_i^{r_i}$ be the maximum order of 
$a_{i1},...,a_{ik_i}$. Define $g:\ZZ_{q_1}^{k_1} \times 
\cdot\cdot\cdot \times 
\ZZ_{q_l}^{k_l} \rightarrow G$ by mapping 
$(x_{11},...,x_{1k_1},...,x_{l1},...,x_{lk_l})$ to $g(\mathbf{x})$ 
$=\prod_{i=1}^l a_{i1}^{x_{i1}}\cdot\cdot\cdot 
a_{ik_i}^{x_{ik_i}}$. Proceed as before.  The only differences are 
that we need to build a huge quantum network to solve the hidden 
subgroup problem and that in computing generators for 
$\ZZ_{q_1}^{k_1} \times\cdot\cdot\cdot\times \ZZ_{q_l}^{k_l}/K$ 
where $K$ is the hidden subgroup $K$ defined by $g$, we need to 
work with a huge matrix when applying Theorem~\ref{reduce} if the 
block structure of the matrix is not exploited. In practice, it is 
therefore desirable to avoid this approach. Furthermore, for each 
prime $p$, instead of using $\ZZ_{q}^k$ where $q = p^r$ is the 
maximum order of the elements $a_j$ (i.e. $r = \max \{t_1, t_2, 
\ldots, t_k\}$ where the order of $a_j$ is $p^{t_j}$), we could use $\ZZ_{p^{t_1}} 
\times 
\ZZ_{p^{t_2}} 
\times 
\ldots \times \ZZ_{p_{t_k}}$. 

 
\bibliographystyle{abbrv} 

\begin{thebibliography}{1} 
 
\bibitem{group} 
J.~A. Beachy and W.~D. Blair. 
\newblock {\em Abstract algebra with a concrete introduction}. 
\newblock Prentice-Hall Inc., New Jersey, 1990. 
 
\bibitem{BL} 
D.~Boneh and R.~J. Lipton. 
\newblock Quantum cryptanalysis of hidden linear functions (extended abstract). 
\newblock {\em Lecture Notes on Computer Science}, 963:424--437. Springer,
Berlin, 1995. 
 
\bibitem{cohen} 
H.~Cohen. 
\newblock {\em Advanced Topics in Computational Number Theory}. 
\newblock Springer-Verlag, New York, 1991. 

\bibitem{Hoy97} 
P.~H{\o}yer. \newblock Conjugated operators in quantum algorithms. 
IMADA preprint, 1997. 
 
\bibitem{KB} 
R.~Kannan and A.~Bachem. 
\newblock Polynomial algorithms for computing the Smith and Hermite normal 
  forms of an integer matrix. 
\newblock {\em SIAM Journal on Computing}, 8(4):499--507, 1979. 
 
\bibitem{mike} 
M.~Mosca. 
\newblock {\em Quantum Computer Algorithms}. 
\newblock D. Phil Thesis, University of Oxford, 1999. 
 
\bibitem{ME} 
M.~Mosca and A.~Ekert. 
\newblock The hidden subgroup problem and eigenvalue estimation on a quantum 
  computer. 
\newblock {\em Lecture Notes in Computer Science}, 1509:174--188, Springer,
Berlin, 1999. 
 
\bibitem{integer} 
A.~Schrijver. 
\newblock {\em Theory of Linear and Integer Programming}. 
\newblock Wiley and Sons, England, 1986. 
 
\bibitem{Sh94} 
P.~W. Shor. 
\newblock Algorithms for quantum computations: Discrete logarithms and 
  factoring. 
\newblock In S.~Goldwasser, editor, {\em Proceedings of the 35th Annual 
  Symposium on Foundations of Computer Science}, pages 124--134. IEEE Computer 
  Society Press, November 1994. 
 
\bibitem{Vaz97} 
U.~Vazirani. 
\newblock UC Berkeley Course CS294-2 Quantum Computation Fall 1997. 
 
\bibitem{Wat00} 
J.~Watrous. 
\newblock Quantum algorithms for solvable groups. 
\newblock preprint, 2000. 
 
\end{thebibliography}
 
\end{document}